\documentclass[aps,onecolumn,floats,rmp]{revtex4}
\usepackage{graphics,graphicx,epsfig}
\usepackage{amssymb,color}
\usepackage{epsf,epstopdf,wrapfig}
\usepackage{amsmath}
\usepackage{natbib}
\usepackage{rotating}

\newcommand{\beq}{\begin{equation}}
\newcommand{\eeq}{\end{equation}}
\newcommand{\beqn}{\begin{eqnarray}}
\newcommand{\eeqn}{\end{eqnarray}}
\newcommand {\e}[1]{\mathrm{~#1}}    
		
\newcommand {\vek}[1]{\mathbf{#1}}
\newcommand{\tr}{{\mbox{\scriptsize {\sc T}}}}

\begin{document}

\title{Learning quadratic receptive fields from neural responses to natural stimuli}

\author{Kanaka Rajan\footnote{krajan@princeton.edu}$^a$, Olivier Marre$^b$ and Ga\v{s}per Tka\v{c}ik\footnote{gtkacik@ist.ac.at}$^{c}$}

\affiliation{$^a$Joseph Henry Laboratories of Physics, \\Lewis--Sigler Institute for Integrative Genomics, Princeton University, Princeton NJ 08544, USA\\
$^b$Institution de la Vision, UPMC UMRS 968, INSERM, CNRS U7210, CHNO Quinze-Vingts, F-75012 Paris, France\\
$^c$Institute of Science and Technology Austria, Am Campus 1, A-3400 Klosterneuburg, Austria}

\date{\today}

\begin{abstract}
Models of neural responses to stimuli with complex spatiotemporal correlation structure often assume that neurons are only selective for a small number of linear projections of a potentially high-dimensional input. Here we explore  recent modeling approaches  where the neural response depends on the quadratic form of the input rather than on its linear projection, that is, the neuron is sensitive to the local covariance structure of the signal preceding the spike. To infer this quadratic dependence in the presence of arbitrary (e.g. naturalistic) stimulus distribution, we review several inference methods, focussing in particular on two information-theory-based approaches (maximization of stimulus energy or of noise entropy) and a likelihood-based approach (Bayesian spike-triggered covariance, extensions of generalized linear models). We analyze the formal connection between the likelihood-based and information-based approaches to show how they lead to consistent inference. We demonstrate the practical feasibility of these procedures by using model neurons responding to  a flickering variance stimulus. 
\end{abstract}

\maketitle

\section{Introduction}
A basic challenge in sensory neuroscience has been to develop a mathematically concise description of how neurons encode stimuli into sequences of spikes. There are two main approaches to this task, which differ primarily in  how much emphasis is placed on anatomical structure versus function. Structure-based modeling starts at the level where basic physical processes govern the observed phenomena. A realistic, conductance-based model could thus be used to predict the neuron's response to a particular type of applied stimulation \cite{hh,koch}. While this bottom-up approach is directly interpretable in terms of biophysical  components and processes, it has a number of disadvantages: (i) the required parameters might be experimentally inaccessible; (ii) in a sensory context, the inputs in this model (the activity of presynaptic neurons) could be related in a complex and intractable manner to the stimulus under experimental control; and, (iii), with enough modeling detail, the problem of understanding or summarizing the ``computation'' that the model implements can  become as difficult as understanding the real neuron itself.

Functional models, in contrast to the above, attempt to capture only the essence of the neural computation: the transformation of stimuli into spiking responses (see \citet{annurev} for an in-depth review). These models are usually fully learned from  data, rather than being derived from the underlying dynamical or physical model  (but see \citet{blaise1, blaise2, hong, lundstrom, ostojic}).  Two considerations are therefore critical to the success of functional models: whether typical electrophysiological recordings can provide enough data for successful inference of the model's parameters; and whether efficient inference algorithms for these parameters exist.  Because the space of all possible stimuli (e.g. all images incident on a retina) and all possible responses (e.g. complete sets of spike arrival times) is vast, our progress must  depend  on making well-chosen simplifying assumptions.  One extreme simplification, for example, involves varying the stimulus along one ``dimension'' only, as in the case of the orientation or wavelength of a drifting grating visual stimulus, and representing the output by a single scalar quantity, e.g. the average firing rate in a chosen time bin. These measurements have traditionally been summarized by tuning curves and have provided  basic insights into principles of sensory and population coding \cite{da}. The relevance of the tuning curve approach is, however, limited by the  choice of the single dimension along which the  stimulus is  manipulated, which may drastically underestimate the true complexity in the structure of the stimuli to which the neuron could respond. Despite strong limitations, such studies helped establish the concept of  a ``receptive field,'' the region of stimulus space where changes in the stimulus modulate the spiking behavior of the neuron.

Central to the concept of receptive field is the notion of locality in the stimulus or feature space. For instance, a ganglion cell in the retina may  be sensitive only to specific changes in light intensity that occur within a small visual angle \cite{rffirst}. A productive way of capturing this notion of locality has been to think of a receptive field as one or more  filters that act on the stimulus; only those stimulus variations that result in the change in filter output have the ability to affect the neural response. In this view, the neurons perform dimensionality reduction by projecting the stimulus down into a small number of dimensions. Consequently, the success of data analysis techniques built around this idea must depend on whether a small number of filters suffices to fully account for the neuron's sensitivity and its response properties. 

\begin{figure*}
\includegraphics[width=4.5in]{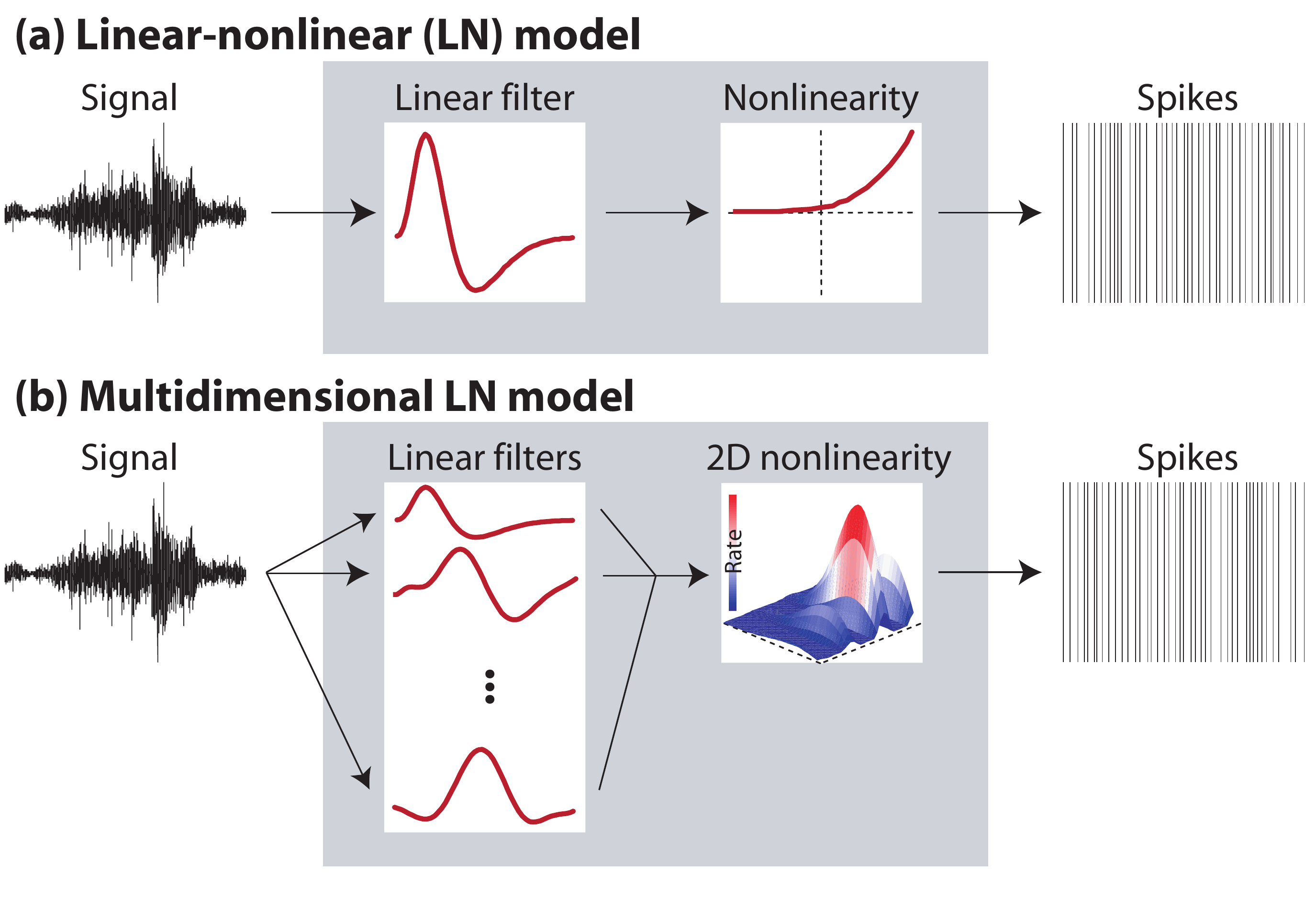}
\centering
\caption{{\bf A schematic showing linear-nonlinear architectures.} {\bf (a)} The instantaneous firing rate, or probability per unit time of emitting a spike, in a linear-nonlinear model neuron is obtained by passing the signal through a linear filter, and mapping the resulting value through a pointwise nonlinearity.  {\bf (b)} A multidimensional LN model neuron requires the signal to be filtered through $K$ linear filters. The number of filters, $K$, is usually much smaller than the dimension of the stimulus. The stimulus projections are mapped into the firing rate through a $K$-dimensional nonlinear function. Without further assumptions, inference of these models is tractable from real recordings only when $K$ is small (usually less than 3).}
\label{cartoon}
\end{figure*}

Methods based in systems identification theory have provided systematic procedures to infer both the receptive fields of neurons as well as subsequent computations. These methods usually  share two key features. First, they can (sometimes necessarily) be used with stimuli that sample the stimulus space broadly, making no explicit assumptions about which stimulus features are important. This is in contrast to the restricted stimuli employed for measuring tuning curves. Second, the procedures usually involve a series of approximations that can provably yield an ever better description of the system if increasing amounts of data are available. Table~\ref{t1} provides an overview of various functional models and related inference methods. Among the earliest to be used successfully, Wiener and Volterra expansions helped identify the first- and second-order kernels mapping the stimulus to response time traces in various systems \cite{wiener,schetzen,jdvknight,recio, marmarelis, sakai}. However, in many cases strong intrinsic nonlinearities attributed to spike generation would require a large number of terms in Wiener-Volterra expansions, despite the fact that the underlying stimulus sensitivity might be simpler, and therefore of low order. Models where the (possibly linear) projections of the stimulus in the receptive field were decoupled from the nonlinearities underlying spiking, as in linear-nonlinear (LN) architectures illustrated in Fig.~\ref{cartoon}, made further progress possible. 

LN and LN-like models have been used widely and profitably to predict the firing rate traces of single sensory neurons, because their parameters can be inferred easily under suitable conditions. However, the more intriguing cases are the ones where LN models either perform poorly or fail entirely. One such failure mode is the inability to account for the statistics of neural activity beyond the mean firing rate. Specifically, real sensory neurons often have variability that is smaller than that attributed to Poisson processes \cite{rrdrbill}; phenomena like refractoriness and spike rate adaptation are not captured by LN models \cite{berrymeister}; and in neural populations, uncoupled LN models fail to reproduce the basic covariance structure of neural activity \cite{schneidman,pillow,granot}. Some of these issues can be addressed by adding suitable dynamical complexity beyond the linear filtering stage, to make the nonlinearities in spike generation more realistic \cite{keat, paninskilif, baccusLNK}, or by including interactions between neurons in models of neural firing \cite{pillow,granot}.  

A different kind of failure of LN models rests on the assumption that stimulus sensitivity occurs through a single (or a small number of) linear projections. One example is contrast adaptation, where a simple LN model derived from a white noise stimulus of a certain variance fails to accurately predict  the response to a stimulus with smaller or larger variance \cite{smirnakis, baccus_meister, rdrvs_mastebroek, borst_egelhaaf, van_hateren}. Other examples include the failure to account for the sensitivity of retinal ganglion cells to fine spatial detail (possibly because of nonlinear summation within the receptive field \cite{sterling}), or to stimulus motion \cite{berry99,schwartz_berry, gollisch_meister}. Generally, these difficulties emerge clearly when the stimulus statistics change or increase in complexity beyond those used to infer the model, for instance by becoming more ``naturalistic''-- i.e. having temporal and spatial pairwise correlation over many scales, skewed first order histograms, and statistical structure beyond second order.

The problems with the LN models can generally be addressed in two possible ways. In the first, LN models can be extended to account for a particular phenomenon on a particular stimulus, e.g., by adding a contrast gain control mechanism \cite{odelia,odelia2} or by an \emph{ad hoc} rescaling of nonlinearities \cite{brenner} to account for contrast adaptation in an experiment where the variance of a Gaussian input is modulated. The second approach is more general: by using complex stimuli, including fully natural movies from the start, the goal is to find the complete (or close to complete) set of features to which the neuron responds. It is worth noting that these two approaches,  as well as the associated use of simple vs natural stimulus ensembles, generally reflect two motivations for building models of neural encoding in the first place: one is to propose and test specific simple models and incrementally improve them, while the other is to infer descriptions that should be valid across a wide range of stimuli and conditions from the start.   For the former purpose---falsifying a model or developing a simple functional form for the stimulus-response relationship---using a stimulus set that is analytically convenient but highly un-natural, e.g., white noise, is sufficient. This is because when a proposed model fails on a subset of stimuli,  it can be excluded or must be extended by additional mechanisms.  Until recently, this was the main reason for using systems identification methods with noise stimuli.  The drive to use naturalistic stimuli comes, on the other hand, from trying to find a model that captures from the start the responses to a wide variety of biologically relevant inputs, and from the observation that naturalistic stimuli may change even the basic filter responses of cells \cite{tanya} and engage response mechanisms that are difficult to probe using noise stimulation (e.g., \citet{olveczky}). Potential  drawbacks with using natural stimuli include technical  obstacles in model inference and the statistical intractability of the natural ensemble \cite{oa,gws}. The choice of the stimulus ensemble  certainly deserves a lengthier discussion; see, for example, \citet{artifice}.

To find the complete set of stimulus features to which a neuron responds, one can look for multiple linear features, a task for which methodological frameworks exist and have been validated for a small number of features. Unfortunately, extracting more than 2 or 3 features becomes intractable because of the curse of dimensionality. A possible anatomically motivated simplification of a multi-feature LN model is a cascade LN (an LNLN) model, where the nonlinearly transformed filter outputs are linearly summed and passed through the spike-generating nonlinearity. Despite some successes \cite{timavh, timiso}, the general problem of inferring cascading models remains technically challenging (usually involving difficult optimizations).  A somewhat simpler LNL system has proven both to account for the behavior of the Y-type retinal ganglion cells very well, as well as being tractable to infer using the sum-of-sinusoids formulation of the Wiener  formalism \cite{jdvknight,vs1,vs2}. A particular case of interest for this review is a special subclass of LNLN models which can be reformulated as  quadratic-nonlinear models, i.e. models where the initial dimensionality reduction of the stimulus is not  a linear projection of the stimulus, but rather an arbitrary quadratic function of the stimulus.

Recently there has been a lot of interest in designing systematic, tractable methods for inferring neural sensitivities when the initial dimensionality reduction step is of high-order (e.g. quadratic)\footnote{When we speak of the order (e.g. linear, quadratic etc), we refer to the order of the kernel operating on the stimulus, which can be defined unambiguously. In contrast, the order of the neural processing system as a whole depends on the stimulus statistics; for example, higher-order statistical structure in the stimulus can conflate first- and second-order responses of the system. Likewise, aspects of the response explained by second order kernel inferred even with Gaussian noise depend on the power spectrum of the input.}. In this paper, we start  by presenting several biologically motivated examples of quadratic stimulus sensitivity in Section~\ref{s1}. We then review several complementary approaches that can be used to learn quadratic stimulus dependence even when neurons are responding to rich, naturalistic stimuli: we discuss the maximally informative stimulus energy \cite{Rajan+Bialek_12} and the maximization of noise entropy \cite{tishby, fitzgerald+al_11, fitzgerald+al_11b} in Section~\ref{s2}, and follow with  the Bayesian spike-triggered covariance \cite{park+pillow_11}  and  related extensions of generalized linear models to quadratic stimulus dependence\footnote{This problem has been worked on by the authors of this review in parallel with the authors of \citet{park+pillow_11}.} in Section~\ref{s3}. We show under which conditions information and likelihood based approaches lead to consistent inference in the Appendix.

%\begin{landscape}
%
\begin{sidewaystable}
\centering
{\scriptsize
\begin{tabular}{|l|l|l|l|}
\hline
{\bf Method} & {\bf Stimulus type} & {\bf Models / restrictions} & {\bf References} \\ \hline\hline
Wiener/Volterra series & white gaussian noise, & $r=r_0 + \mathbf{k}\cdot \mathbf{s} + \mathbf{s}^T\mathbf{Q}\mathbf{s} + \cdots $ & \tiny{\cite{wiener,marmarelis,schetzen}} \\
& sum-of-sinusoids & &\tiny{\cite{jdvknight,recio}} \\\hline 
spike trigger average (STA) & spherically symmetric, & LN (single filter), isolated spikes & \tiny{\cite{deboer,mseq,paninski03}} \\ 
(reverse correlation) & binary noise, m-sequences & $r=f(\mathbf{k}\cdot\mathbf{s})$ & \tiny{ \cite{simoncelli04,stnc}} \\\hline
debiased STA & ``gaussian-like'' & LN (single filter), isolated spikes & \tiny{\cite{lesica} }\\ 
(reverse correlation) & asym. 1-point histogram& $r=f(\mathbf{k}\cdot\mathbf{s})$ &  \\\hline
spike trigger covariance (STC) & gaussian & LN (multiple filters), isolated spikes & \tiny{\cite{stc,fnd}}  \\  
(reverse correlation) & & $r=f(\mathbf{k}_1\cdot \mathbf{s}, \dots,\mathbf{k}_K\cdot\mathbf{s})$ & \tiny{\cite{simoncelli04,fairhall06,maravall,odelia}} \\ \hline
extended projection pursuit regression (ePPR) & any & LN (multiple filters) & \tiny{\cite{rapela}}  \\  
& & $r=f(\mathbf{k}_1\cdot \mathbf{s}, \dots,\mathbf{k}_K\cdot\mathbf{s})$ &  \\ \hline
iSTAC & gaussian & LN (multiple filters) & \tiny{\cite{istac}}\\ 
(reverse correlation)& & $r=f(\mathbf{k}_1\cdot \mathbf{s}, \dots,\mathbf{k}_K\cdot\mathbf{s})$ &  \\ \hline
differential reverse correlation (dRC) & spike triggering snippet &linear feature that predicts spike timing & \tiny{\cite{gtmm}} \\
(reverse correlation) &&$t_{spike}\propto \mathbf{k}\cdot\mathbf{s}$& \\ \hline
maximally informative dimensions (MID) & any & LN (multiple filters)  & \tiny{\cite{mid,tanya,Kouh}} \\ 
(info maximization) & &$r=f(\mathbf{k}_1\cdot \mathbf{s}, \dots,\mathbf{k}_K\cdot\mathbf{s})$ & \\\hline
(maximum likelihood) & any & leaky integrate and fire (LIF/LN-LIF) & \tiny{\cite{gerstner,paninskilif,pillow2}}  \\ \hline
error function minimization &  & dynamical extensions of LN& \\ 
(general fitting methods) & any & $r=\Theta(h)\dot{h},h=\mathbf{k}\cdot\mathbf{s}+\mathbf{q}\cdot\mathbf{y}+\eta$ (Keat),  & \tiny{\cite{keat,baccusLNK}}  \\
&& $\dot{A}_i=M_{ij}(f(\mathbf{k}\cdot\mathbf{s}))A_j,r=A_1$ (LNK) &\\ \hline
generalized linear models (GLM) & any & point process (dependence on past spiking) & \tiny{\cite{gln1,gln2,pillow}} \\
(maximum likelihood)& & $r=f(\mathbf{k}\cdot\mathbf{s} + \mathbf{q}\cdot\mathbf{y} + $ (effect of other neurons) $)$ & \tiny{\cite{pillow2,gerwinn}} \\ \hline
isoresponse mapping & synthetic stimuli & LNLN cascade & \tiny{\cite{timavh,timiso}} \\
& (parametrizable, low-D) & $r=f(\mathbf{k}_1\ast g(\mathbf{k}_2\ast \mathbf{s}))$ &  \\ \hline
maximally informative stim. energy (MISE) & any & general quadratic model  &\tiny{\cite{Rajan+Bialek_12}}  \\ 
(info maximization) & & $r=f(\mathbf{k}\cdot\mathbf{s},\mathbf{s}^T\mathbf{Q}\mathbf{s})$ & \\ \hline
maximization of noise entropy & any & $r=\mathrm{logistic}(k_0+\mathbf{k}\cdot\mathbf{s} + \mathbf{s}^T\mathbf{Q}\mathbf{s})$ & \tiny{\cite{tishby,fitzgerald+al_11,fitzgerald+al_11b}} \\
(convex optimization)& & & \\ \hline
Bayesian STC / quadratic GLM & any & additive linear and quadratic contributions & \tiny{\cite{park+pillow_11}}\\ 
(likelihood maximization) & & $r=f(\mathbf{k}\cdot\mathbf{s} + \mathbf{s}^T\mathbf{Q}\mathbf{s})$ & \\ \hline
\end{tabular}
}
\caption{{\bf Functional models for single neurons and the related inference methods.} $r(t)$ is the firing rate or the probability of spiking; $\mathbf{k}$ are linear filters acting on stimulus clips $\mathbf{s}$; $\mathbf{Q}$ is a quadratic kernel (any symmetric matrix); $\mathbf{q}$ is a linear filter on the sequence of past spikes $\mathbf{y}$; $f,g$ are  arbitrary nonlinear functions; $\ast$ denotes a convolution; $\Theta$ is a thresholding operation (1 when the argument crosses some threshold from below, 0 otherwise); $\eta$ is a white noise Langevin force.  In this paper, we use the term "gaussian" to denote stimuli whose components are jointly Gaussian and possibly correlated (i.e. non-white), unless otherwise stated. While reverse correlation methods are formally simpler for uncorrelated (white) Gaussian noise, it is possible to generalize them for use with correlated noise ensembles. For example, to compute an unbiased estimate of the linear (L) part of the model using STA and a correlated stimulus, one needs to correct for stimulus correlations by acting on the spike triggered average with the inverse covariance matrix.  For an extensive review of spike-triggered (reverse correlation) methods, see \citet{stnc}.  }
\label{t1}
\end{sidewaystable}
%
%\end{landscape}
%
%
\section{High-order stimulus dependence}
\label{s1}

In a typical experiment, a neuron can be driven by a synthetically generated stimulus containing a desired statistical structure. For probing the visual system for example, this stimulus might be a random binary checkerboard, a drifting grating, or full-field light intensity flicker. If the neuron's response depends solely on the stimulus presented in the recent past of duration $T$ (and possibly on its own previous spiking behavior), we can restrict our attention to stimulus clips $\vek{s}$ of length $\geq T$. These clips are drawn from a distribution $P(\vek{s})$ that characterizes the stimulus; the $N$ components of vector $\vek{s}$  represent successive stimulus values in time and optionally across space.  Our task is then to infer the dependence of the instantaneous probability of spiking  (firing rate) at time $t$ on the stimulus, $\mathbf{s}(t)$, presented just prior to $t$.
  
 If the neuron is well described by the linear-nonlinear (LN) model,  where the spiking rate $r$ is an arbitrary positive, point-wise, nonlinear function $f$ of the stimulus projected onto the filter, $r(\mathbf{s})=f(\vek{k}\cdot\vek{s})$, and the stimulus distribution is chosen to be spherically symmetric, $P(\vek{s})=P(|\vek{s}|)$, we can use the spike-triggered average (STA) to obtain an unbiased estimate of the single linear filter $\vek{k}$ \cite{deboer,simoncelli04}. Spike-triggered covariance (STC) generalizes the filter inference to cases where the firing rate depends nonlinearly on $K\geq 1$ projections of the stimulus, $r(\vek{s})=f(\vek{k}_1\cdot\vek{s},\vek{k}_2\cdot\vek{s},\dots,\vek{k}_K\cdot\vek{s})$ \cite{stc}. The number of relevant linear filters, $K$, is equal to the number of nonzero eigenvalues of the spike-triggered covariance matrix. A successful application of STC requires $P(\vek{s})$ to be Gaussian, and the number of filters $K$ be small (usually $\leq 3$) to ensure an adequate sampling of the filters and the nonlinearity $f$, given the data obtained in the typical experiment (however when inferring only the linear part of such models as many as $14$ filters have been estimated \cite{rust}). STC has been used successfully, for example, to understand the computations performed by motion sensitive neurons in the blowfly \cite{fnd}, to map out the sensitivity to full-field flickering stimuli  in salamander retinal ganglion cells \cite{fairhall06},  to explore contrast gain control \cite{odelia,rust}, and to understand adaptation in the rodent barrel cortex \cite{maravall}.

Before moving on, it seems appropriate to return once more to the Wiener formalism and contrast it with  spike-triggered methods for recovering LN models. The underlying assumptions of the two approaches may seem substantially different: first, because of the presence of the nonlinear (N) transformation in the LN model, and second, because the output of the LN model is usually taken to predict the rate of a stochastic point process, while Wiener series is intended for analyzing deterministic systems \cite{wiener}. Nevertheless, it is easy to see that when uncorrelated (i.e. \emph{white}) Gaussian noise is used to extract the filters of the LN model using spike triggered average (STA) and spike triggered covariance (STC), STA and STC also provide unbiased estimates (up to a scaling factor) of first- and second-order Wiener kernels. The difference arises in subsequent analysis steps: in  case of LN models, STA and STC are used solely as dimensionality reduction steps to identify the relevant subspace of the stimuli in which the nonlinear transformation acts, while in the Wiener formalism, STA and STC literally are the first two terms in a functional expansion that provides the best least-squares fit to the observed firing rate. \citet{vj} have further demonstrated that the Wiener formalism is a special case of a  general probabilistic maximum entropy framework for describing joint distributions of stimuli and responses. In this framework, for example, the classic Wiener formalism is recovered if the stimulus distribution is Gaussian, and the response variable is also Gaussian with additive noise. If, on the other hand, the output variable is binary (spike / no-spike), the same maximum entropy approach reduces to identifying LN-type models with exponential nonlinearities.

While powerful and simple to use, spike-triggered covariance (STC) only works if Gaussian stimuli are employed, and is feasible only if $K$ is small. The Gaussian ensemble can be a serious restriction for neurons that do not respond well (or at all) to unstructured stimuli; furthermore, we are likely to miss several neural mechanisms that depend on naturalistic statistical structure, such as correlations, intermittency etc, if the neuron responds to Gaussian stimulation. A versatile method should therefore be able to successfully infer the multiple-filter dependence of a neuron probed with a stimulus of arbitrary complexity. Maximally informative dimensions (MID) \cite{mid} or likelihood inference for single-filter generalized linear models \cite{gln1, gln2, pillow, pillow2, gerwinn} have been used to this end when the dependence is linear, but the attempts to incorporate full quadratic stimulus dependence have been less common. 

There are several instances of quadratic stimulus dependence. Let us  consider  a situation where the neuron has a vanishing spike-triggered average, as with a complex cell, non--phase--locked auditory neurons \cite{recio}, or motion-sensitive neurons. In these cases a natural starting point would be a search for more than a single linear filter. For a model complex cell in the visual cortex, we would find two phase-shifted vectors $\vek{k}_1$ and $\vek{k}_2$ that together form a quadrature pair, such that the most informative variable concerning the neuron's firing is the ``power,''
\begin{equation}
r(\vek{s})=f\left[ (\vek{k}_1\cdot\vek{s})^2+ (\vek{k}_2\cdot\vek{s})^2\right].\label{cc}
\end{equation}

\begin{figure*}
\includegraphics[width=\textwidth]{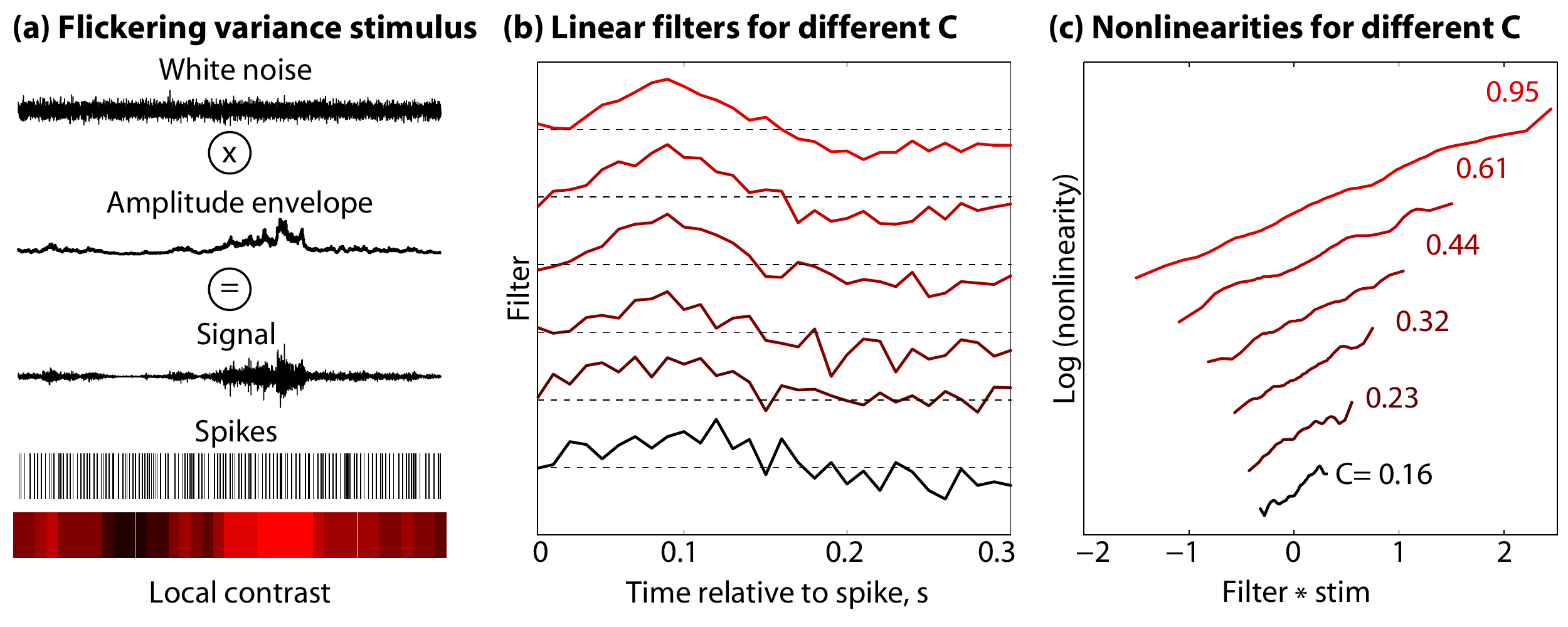}
\centering
\caption{{\bf A synthetic contrast-adapting neuron probed with the ``flickering variance'' stimulus.} The instantaneous spiking rate is given by $r(t) = f(\mathbf{k}_0\cdot \mathbf{s}(t) + \mathbf{s}(t)^T\mathbf{Q}\mathbf{s}(t)+\mu)$, where $f(\cdot)=\log(1+\exp(\cdot))$, $\mu$ is an offset (bias), and the quadratic kernel $\mathbf{Q}$ is a rank 2 matrix with a quadrature eigenvector pair. {\bf (a)} The stimulus is sampled at $\Delta=1\e{ms}$ scale and is given by $s(t) = \exp(A(t)) w(t)$, where $w(t)$ is given by uncorrelated white noise of fixed variance, and $A(t)$ is a gaussian noise process with correlation time $\tau_c=1\e{s}$. The stimulus can be chopped into segments of duration $\tau\leq \tau_c$, which can be sorted by local contrast $C$ (intensity of red). Spike-triggered average analysis can be applied to recover effective LN models for all stimulus segments sharing the same local contrast. {\bf (b)} The linear filters recovered at various contrast levels (shade of red; filters displaced along vertical axis for easier readability). At lower contrasts the neuron produces less spikes, making the filter estimate more noisy, but the filter shape is constant across a range of $C$ and closely approximates the model filter $\mathbf{k}_0$. {\bf (c)} The nonlinearities for different contrast levels $C$ (plot legend, shades of red; the nonlinearities displaced along vertical axis for easier readability). The slope of the nonlinearity decreases with increasing contrast (although the adaptation is not perfect, in this example), to prevent quick saturation of the response at high $C$. }
\label{fa}
\end{figure*}

Similarly, models of contrast gain control in the retina also include sensitivity to second-order features in the stimulus, with the spiking probability of the form \cite{odelia},
\begin{equation}
r(\vek{s})=\frac{f(\vek{k_0}\cdot \vek{s})}{\sum_{i=1}^M w_i(\vek{k}_i\cdot\vek{s})^2+\sigma^2},\label{gainctrl}
\end{equation}
where the quadratic terms in the denominator scale down the gain at high contrast (in this case however, the neuron has a non-vanishing linear filter $\vek{k}_0$). A simulated model neuron showing contrast adaptation is shown in Fig.~\ref{fa}a, featuring both the first- and second-order stimulus sensitivity. The model neuron is probed with a ``flickering variance'' stimulus, in which the variance of the white noise (with a very short correlation time) is dynamically modulated by a noise process correlated across a longer timescale (c.f.~\citet{fnature}). With this synthetic stimulus, the separation of timescales allows us to partition the stimulus into chunks with approximately constant variance in luminance, $\sigma_L^2$. This variance is directly related to the temporal contrast, $C=\sigma_L/\bar{L}$, because the average mean light intensity $\bar{L}$ is kept constant. Within each stimulus segment, we can use STA to recover the LN model, as shown in Figs.~\ref{fa}b,c. Our real goal, however, is to infer a joint model valid across the whole stimulus, and to do so ultimately with naturalistic stimuli with scale-free power spectra, where no clear separation exists between the fast fluctuation and slow variance modulation.

We can describe these and similar examples by a generic ``quadratic'' model neuron which is sensitive to a second-order function of the input (parametrized by a real, symmetric matrix $\vek{Q}$) in addition to the linear projection (parametrized by the filter $\vek{k}_0$):
\begin{equation}
r(\vek{s})=f(\vek{k}_0\cdot \vek{s},\;\; \vek{s}^\tr \vek{Q}  \vek{s}).	\label{e1}
\end{equation}
Graphically, while a threshold LN model with a linear filter corresponds to a classifier whose separating hyperplane is perpendicular to the filter, the proposed LN model with a threshold nonlinearity and a quadratic filter $\vek{Q}$ is selective for all stimuli that lie outside an $N$-dimensional ellipsoid whose axes correspond to the eigenvectors of $\vek{Q}$. 

For the contrast gain control model described in  Eq~(\ref{gainctrl}) the matrix $\vek{Q}$ is of rank $M$, with eigenvalues $w_i$ and eigenvectors $\vek{k}_i,i>0$.  The complex cell example described in  Eq~(\ref{cc}) has $\vek{k}_0=0$ and $\vek{Q}=\sum_{i=1}^2 \vek{k}_i \vek{k}_i^\tr$; in other words, $\vek{Q}$ is a rank $2$ matrix. While these examples feature quadratic dependences involving matrices of low rank, it is possible to extend these models to biologically relevant cases where the matrix does not have to be low rank \cite{Rajan+Bialek_12}. For example, the probability of spiking could be a nonlinear function of the ``power'' $p(t)$, $r(t)=f[p(t)]$, where the power is given by:
\begin{equation}
p(t) = \int d\tau f_2(\tau) \left[\int dt' f_1(t-\tau-t')s(t')\right]^2;
\end{equation}
here $s(t)$ is the stimulus, and $f_1$ and $f_2$ are linear filters, such as those used to describe non-phase-locked auditory neurons. If the smoothing time of the second filter $f_2$ is larger than that of the first filter $f_1$, it has been shown in \cite{Rajan+Bialek_12} that the quadratic kernel $\mathbf{Q}$ for this model has a rich (full-rank) spectrum.

In the next section we review methods that permit inference of low- or full-rank quadratic kernels, $\mathbf{Q}$.

\section{Inferring quadratic stimulus dependence from data}
Every real, symmetric matrix can be spectrally decomposed into $\vek{Q}=\sum_{i=1}^N \lambda_i \vek{k}_i\vek{k}_i^\tr$. The response of the quadratic model is thus $r=f\left[\sum_{i=1}^N \lambda_i (\mathbf{k}_i\cdot\mathbf{s})^2\right]$, explicitly demonstrating that quadratic models are special cases of the LNLN cascade, where the first linear stage involves applying the filters $\mathbf{k}_i$, the first nonlinear stage squares the projections, the second linear stage is a summation with weights $\lambda_i$, and the last nonlinear transformation is $f(\cdot)$.  The spectral decomposition implies that we could try recovering the quadratic dependence of $\vek{Q}$ in Eq~(\ref{e1}) by, for example, multidimensional MID (see Table~\ref{t1}), hoping to infer all $\{\vek{k}_i\}$ as orthogonal informative dimensions. While formally true, this is infeasible in practice because maximizing the mutual information would involve sampling $N$-dimensional distributions from stimulus samples that are limited in number by the number of spikes \cite{mid}. The same sampling problem would reappear when trying to estimate the nonlinearity, $f(\mathbf{k}_1\cdot\mathbf{s},\mathbf{k}_2\cdot\mathbf{s},\dots,\mathbf{k}_N\cdot\mathbf{s})$. 

To address this problem efficiently, we formulate the inference problem by explicitly assuming quadratic dependence on the stimulus: in this case, the stimulus immediately gets projected down to a single scalar variable $x=\vek{s}^\tr \vek{Q}\vek{s}$, meaning that information-theoretic quantities, the likelihood, as well as the nonlinearity $f(\cdot)$ will only depend on the stimulus through $x$. This makes  inference problem tractable even when $\vek{Q}$ is of  high rank. Clearly, this advantage is gained by assuming that  projections onto eigenvectors of $\vek{Q}$ combine as a sum of squares. This assumption is not a mere mathematical convenience:  as we have shown previously,  well-known phenomena of phase invariance, adaptation to local contrast or sensitivity to the signal envelope are all examples of true second-order stimulus sensitivity in real neurons. Additionally, response phenomena in the visual cortex grouped together as relating to the \emph{non-classical receptive field} could also be manifestations of quadratic or higher-order sensitivity \cite{zetzsche}. 

\subsection{Finding  quadratic filters using information maximization}
\label{s2}
Despite their utility and simplicity, spike-triggered methods require the use of statistically simple stimuli and in particular, exclude the use of stimuli with naturalistic statistics, e.g. those with $1/f$ spectra, non-Gaussian histograms and/or high-order correlations. This is a big challenge when studying neurons beyond the sensory periphery that are responsible for extracting high-order structure, or neurons unresponsive to white noise presentations, for example those in the auditory pathway. To address this issue and recover the filter(s) in an unbiased manner with an arbitrary stimulus distribution, maximally informative dimensions (MID) \cite{mid, tanya, Kouh} have been developed and utilized to recover simple cell receptive fields, among other examples. MID looks for a linear filter $\vek{k}$ that maximizes the information between the presence/absence of a spike and the projection $x$ of the stimulus onto $\vek{k}$, $x=\vek{k}\cdot\vek{s}$. Information per spike is then given by the Kullback-Leibler divergence of $P(x|\mathrm{spike})$, the \emph{spike-triggered distribution} (the distribution of stimulus projections preceding the spike) and $P(x)$, the \emph{prior distribution} (the overall distribution of projections): 
\begin{equation}
I_{\rm spike}=D_{KL}\left[P(x|\mathrm{spike}) || P(x)\right]=\int dx\;P(x|\mathrm{spike})\log_2\frac{P(x|\mathrm{spike})}{P(x)}. \label{iopt}
\end{equation}
Given the spike train and the stimulus, finding  $\vek{k}$ becomes an information optimization problem in $I_{\rm spike}$ that can be solved using various annealing methods, although the existence of local extrema makes this a nontrivial task. 

Spike-triggered methods and MID do not explicitly assume a form for the nonlinearity $f(\cdot)$ in the LN model; instead, they provide unbiased estimates of the filter(s), and once the filters are known, the nonlinearity can be reconstructed using the Bayes' rule from sampled spike-triggered and prior distributions:
\begin{equation}
f(x)\propto P(\mathrm{spike}|x)=\frac{P(x|\mathrm{spike}) P(\mathrm{spike})}{P(x)}, \label{nonlin}
\end{equation}
where $ P(\mathrm{spike})$ is directly proportional to the average firing rate during the experiment. 

In classical MID, one finds a (set of) linear filter(s) by maximizing Eq.~(\ref{iopt}) with respect to $\mathbf{k}$. In~\citet{Rajan+Bialek_12}, this approach was extended to quadratic stimulus sensitivity, as follows. A quadratic filter $\vek{Q}$  can  be reconstructed from an observed spike train by maximizing the information in Eq~(\ref{iopt}), where $x$ is now given by $x=\vek{s}^\tr\vek{Q}\vek{s}$. Taking a derivative of Eq~(\ref{iopt})  with respect to $\vek{Q}$ gives us a gradient,
\begin{equation}
\nabla_{\mathbf Q} I = \int\!\! dx\;P_{\vek{Q}}(x)\left[\left\langle\left.\mathbf s \mathbf s^\tr \right|x, \mathrm{spike}\right\rangle-\left\langle\left.\mathbf s \mathbf s^\tr \right|x\right\rangle\right]\frac{d}{dx}\left(\frac{P_{\mathbf Q}(x|\mathrm{spike})}{P_{\mathbf Q}(x)}\right), \label{ideriv}
\end{equation}
where $\langle\;\cdot\;\rangle$ indicates averaging over the spike-triggered and prior distributions respectively, and the subscript $\vek{Q}$ makes the dependence of the probability distributions explicit. Only the symmetric part of $\vek{Q}$ contributes to $x$, and the overall scale of the matrix is irrelevant to the information, making the number of free parameters $N(N+1)/2-1$. 

To learn the ``Maximally Informative Stimulus Energy'' or the quadratic filter $\vek{Q}$, we can ascend the gradient in successive learning steps \cite{Rajan+Bialek_12},
\begin{equation}
\mathbf Q \rightarrow \mathbf Q+\gamma\;\nabla_{Q} I. \label{lstep}
\end{equation}
The probability distributions within the gradient are obtained by computing $x$ for all stimuli, choosing an appropriate binning for the variable $x$, and sampling binned versions of the spike-triggered and prior distributions. The $\langle \vek{s}\mathbf s^\tr\rangle $ averages  are computed separately for each bin; and the integral in Eqs~(\ref{iopt},\ref{ideriv}) and the derivative in Eq~(\ref{ideriv}) are approximated as a sum over bins and as a finite difference, respectively. To deal with local maxima in the objective function, we use a  large starting value of $\gamma$ and gradually decrease $\gamma$ during learning. This basic algorithm can be extended by using kernel density estimation and stochastic gradient ascent/annealing methods, but we do not report these technical improvements here. 

It is possible to select an approximate linear basis in which to expand the matrix $\vek{Q}$, by writing
\begin{equation}
\mathbf{Q} = \sum_{\mu=1}^M \alpha_\mu\mathbf{B}^{(\mu)}. \label{basis}
\end{equation}
The basis can be chosen so that increasing the number of basis components $M$  allows the reconstruction of progressively finer features in $\vek{Q}$. We considered as $\{\vek{B}^{(\mu)}\}$ a family of Gaussian bumps that tile the space of the $N\times N$ matrix $\vek{Q}$ and whose scale (standard deviation) is inversely proportional to $\sqrt{M}$. For $M\rightarrow N^2/2$ the matrix set becomes a complete basis, allowing every $\vek{Q}$ to be exactly represented by the vector of coefficients $\vek{\alpha}$. In such a matrix basis representation, the learning rule becomes
\begin{equation}
\mathbf \alpha_\mu \rightarrow \mathbf \alpha_\mu+\gamma\sum_{i,j=1}^N\frac{\partial I}{\partial \mathbf{Q}_{ij}}\mathbf{B}_{ij}^{(\mu)}, 
\end{equation}
where applying the chain rule on $\nabla_\mathbf QI$ yields the $\mathrm{Trace}[\nabla_\mathbf Q(\alpha)\cdot\mathbf B]$ update term at each step. 

\begin{figure*}
\includegraphics[width=\textwidth]{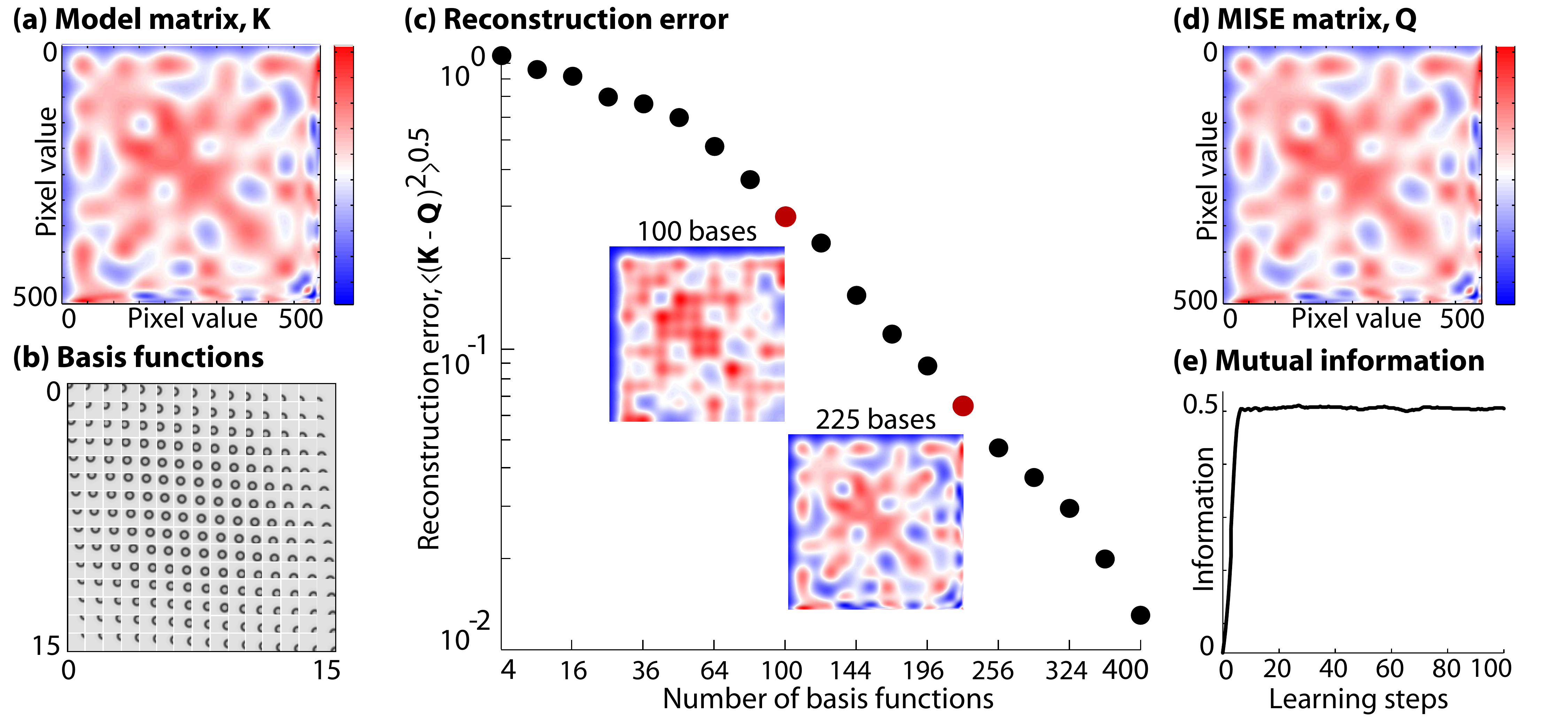}
\centering
\caption{{\bf Reconstructing a high rank quadratic filter using stimuli extracted from natural scenes.} {\bf (a)} A complex  high-rank randomly generated matrix ${\mathbf K}$ will be used as a quadratic filter of a model cell that fires whenever $\mathbf{s}^T\mathbf{K}\mathbf{s}$ exceeds a fixed threshold. $\mathbf{K}$ is thus the true quadratic filter for our threshold LN model neuron.  {\bf(b)}  A collection of $225$ Gaussian matrix basis functions whose peaks densely tile the matrix space; a trial matrix is constructed as a linear sum (with coefficients $\{\alpha_\mu\}$) of the basis matrices, and information optimization is performed over  $\{\alpha_\mu\}$. {\bf (c)}  The normalized reconstruction error, shown in black dots, decreases as the number of basis functions $M$ increases from $4$ to $400$; with enough data perfect reconstruction is possible as $M$ approaches the number of independent pixels in $\vek{K}$. The two red dots show reconstructions with $M=100$ or $M=225$ basis functions, respectively.
{\bf(d)} The reconstructed, maximally informative matrix kernel $\vek Q$ after maximizing mutual information using $400$ basis functions. {\bf(e)} Mutual information increases as learning progresses in steps given by Eq.~(\ref{lstep}), peaks at step $40$ and remains unchanged thereafter. Learning step $100$ is the point where the maximally informative $\vek Q$ is extracted.}
\label{f3}
\end{figure*}

We illustrate this approach with two examples. In the first example we make use of the matrix basis expansion from Eq~(\ref{basis}) to infer a quadratic kernel $\vek{K}$ that is of arbitrarily high rank. For $\mathbf {K}$ we used a highly-structured $500\times500$ matrix as shown in Fig.~\ref{f3}(a). While this is not an example of a receptive field from a real neuron, it illustrates the validity of the approach even when the response has an atypical and highly structured dependence on the stimulus. The stimuli were natural image clips from the Penn Natural Image database, flattened into a high-dimensional vector representation $\mathbf{s}$ \cite{pidb}, and the spikes were generated by thresholding the term $\mathbf {s^\tr Ks}$. Gaussian basis matrices, similar to the $225$ shown in Fig.~\ref{f3}(b) were used to expand the quadratic kernel, reducing the number of optimization parameters from $\sim 2.5\times10^5$ to a few hundred. We start the gradient ascent with a large $\gamma$ value of $1$ and progressively scale it down to $0.1$ near the end of the algorithm;  Fig.~\ref{f3}(e) shows the information plateauing in about $20$ learning steps. The maximally informative quadratic filter reconstructed from $400$ basis coefficients is shown in Fig.~\ref{f3}(d). Figure~\ref{f3}(c) demonstrates how the root-mean-squared reconstruction error systematically decreases as the number of basis functions $M$ is increased from $4$ to $400$, improving precision. Insets show $2$ inferred matrices with $M =100$ (corresponding to the first dot) showing a marked improvement  with $M=225$ (corresponding to the second red dot). Reconstruction error drops to $\sim1\%$ for $M=400$.

In contrast to standard MID where the number of spikes required grows exponentially in the number of filters extracted, the data requirement for this approach is proportional to the square of the stimulus dimension for a matrix kernel with no additional structural simplifications (these data requirement- and performance-related issues are explored in detail in~\citet{Rajan+Bialek_12}). For the examples shown in the paper, expansion in matrix basis reduces this number to the order of stimulus dimension, making this procedure  pertinent for experimentalists. 

The second example shows the MISE analysis of the synthetic neuron presented in Fig.~\ref{fa} where stimulus-response relationship is more biologically realistic, through a smooth nonlinear function $f$ and both a linear as well as a quadratic kernel. The analysis is applied to the flickering variance stimulus without partitioning it into regions of fixed contrast. With $\sim 2\times 10^4$ spikes, the method recovers the linear filter $\mathbf{k}_0$ as well as the quadratic kernel, which turns out to have the two dominant eigenvectors $\mathbf{k}_1,\mathbf{k}_2$, corresponding to the  quadrature pair of filters used to construct $\mathbf{Q}$, as shown in Fig.~\ref{fb}b.

These examples show that quadratic filters can be extracted using information maximization for both low-rank and full-rank matrices, under natural stimulation and with a realistic numbers of spikes. Importantly, for cases where the stimulus sensitivity is both linear and quadratic, MISE does not explicitly assume that the effects of two filtering operations are additive, i.e. that $x=\mathbf{k}_0\cdot\mathbf{s} + \mathbf{s}^\tr\mathbf{Q}\mathbf{s}$; rather, the dependence can be an arbitrary 2D nonlinear function, $f(\mathbf{k}_0\cdot\mathbf{s},\mathbf{s}^\tr\mathbf{Q}\mathbf{s})$. Unlike the quadratic generalizations of GLM presented below, this allows MISE to fully recover forms of contrast gain control that have a  parametric form similar to Eq.~(\ref{gainctrl}).

\subsection{Finding quadratic filters using maximization of noise entropy}
Another information-theoretic approach for inferring single neuron sensitivities is derived from the principle of noise entropy maximization \cite{tishby, fitzgerald+al_11, fitzgerald+al_11b}.
Suppose that the spiking or silence of a chosen neuron in a time bin indexed by $t$ is represented by a binary variable $y_t\in \{0,1\}$. From data, we can reliably estimate certain statistics of the neural response, such as the average spiking rate $\langle y_t\rangle_t$, the spike-triggered average $\langle y_t\mathbf{s}(t)\rangle_t$, or the spike-triggered covariance $\langle y_t\mathbf{s}(t)\mathbf{s}(t)^\tr\rangle_t$, where the brackets $\langle\;\cdot\;\rangle_t$ denote averaging across the duration of the experiment. In general, all these statistics are of the form $\langle O_\mu(\mathbf{s})y_t\rangle_t$, where $\mu$ indexes the different operators whose expectation values we are computing. 

The crucial step is to look for maximum entropy approximations to $P(y|\mathbf{s})$, the distribution of the (binary) neural response given the stimulus. Maximum entropy distributions  are as unstructured (random, therefore parsimonious) as possible with the constraint that they exactly reproduce the measured expectation values of a chosen set of  statistics, $\{O_\mu\}$ \cite{jaynes1,jaynes2}. When the variable $y$ is binary, it can easily be shown that these distributions have the form of the logistic function,
\begin{equation}
P(y=1|\mathbf{s}) = \frac{1}{1+e^{-F(\mathbf{s})}}, \label{nent}
\end{equation}
where $F$ resembles the free energy in statistical physics:
\begin{equation}
F(\mathbf{s})=\sum_\mu \lambda_\mu O_\mu(\mathbf{s}),
\end{equation}
and $\lambda_\mu$ are the Lagrange multipliers that have to be set such that the set of statistics measured in the data equals the expectation values of the same operators under distribution $P$, i.e. $\langle O_\mu(\mathbf{s})y\rangle_P = \langle O_\mu(\mathbf{s})y\rangle_t$. To apply this general framework to the inference of quadratic filters, the authors of~\citet{fitzgerald+al_11b} choose the mean firing rate, STA and STC as constraints, which yields the following response distribution:
\begin{equation}
P(y=1|\mathbf{s})= \frac{1}{1+\exp(\mu + \mathbf{k}_0 \cdot \mathbf{s} + \mathbf{s}^\tr \mathbf{Q}\mathbf{s})},
\end{equation}
where $\{\mu,\mathbf{k}_0,\mathbf{Q}\}$ act as the Lagrange multipliers $\lambda_\mu$ conjugated to the operators $\{y,y\mathbf{s},y\mathbf{s}\mathbf{s}^\tr\}$. Numerically, the task is to solve for parameters $\{\mu,\mathbf{k}_0,\mathbf{Q}\}$ that satisfy a set of constraints: $\langle y\rangle_t = \langle y \rangle_P$ (matching the measured mean firing rate to that of the model), $\langle y\mathbf{s}\rangle_t = \langle y \mathbf{s}\rangle_P$ (matching the measured STA to that of the model), and $\langle y\mathbf{s}\mathbf{s}^\tr\rangle_t = \langle y\mathbf{s}\mathbf{s}^\tr \rangle_P$ (matching the measured STC to that of the model). This is a convex optimization task and can be solved by conjugate gradient descent. 

An attractive feature of this approach emerges when we rewrite the information per spike $I(\mathrm{spike};\mathbf{s})$ as a difference between the total and the noise entropy as follows:
\begin{equation}
I(\mathrm{spike};\mathbf{s})= \sum_{\mathbf{s}} P(\mathbf{s}) \sum_yP(y|\mathbf{s})\log_2\frac{P(y|\mathbf{s})}{P(y)}=S[P(y)] - \langle S[P(y|\mathrm{s})]\rangle_\mathbf{s},
\end{equation}
where $S[P(x)]=-\sum_x P(x)\log_2P(x)$ is the entropy of $P(x)$. The first term (total entropy) is fully determined by the mean spiking rate $\langle y \rangle_t$, $S[P(y)]=-\langle y\rangle_t \log_2\langle y\rangle_t  - (1-\langle y\rangle_t ) \log_2(1-\langle y\rangle_t )$ because $y$ is a binary variable. The mean firing rate is one of the statistics constrained in the model for $P(y|\mathbf{s})$, ensuring consistency. Since our model for $P(y|\mathbf{s})$ has maximum entropy given the observed constraints, we are effectively setting an upper bound on the noise entropy $\langle S[P(y|\mathbf{s})]\rangle_\mathbf{s}$, and therefore a lower bound on the mutual information $I$. As more and more statistics $O(\mathbf{s})$ are included as constraints into the maximum entropy model for Eq.~(\ref{nent}), the noise entropy must progressively drop and information increase towards the true value (which is bounded by the output entropy). At the point where this lower bound on information meets the actual information per spike (which can be empirically estimated from, e.g., repeated stimulation \cite{brenner}), we obtain the complete list of the relevant stimulus statistics $\{O_\mu\}$ that characterize the sensitivity of the neuron.

In~\citet{fitzgerald+al_11b}, the authors show that this framework is applicable for inferring quadratic neural filters on synthetic and real data, and compare it to MID. This method is applicable to any stimulus ensemble, but requires  assumptions beyond those needed for MID or MISE: namely, that the nonlinear function is logistic, and that the contributions of the linear and quadratic filters add. The advantage of the method is that the problem is convex, does not suffer from the exponential curse of dimensionality (like multi-dimensional MID), and is flexible, permitting various new constraints (beyond the STA and STC) to be used in constructing models for the stimulus-conditional distribution $P(y|\mathbf{s})$.
\subsection{Finding quadratic filters in a likelihood framework: GLM extensions and Bayesian STC}
\label{s3}
A powerful technique for modeling neural spiking behavior is  the generalized linear model (GLM) framework \cite{gln1,gln2}. Recently GLM has been used to account for  the stimulus sensitivity, dependence on spiking history, and connectivity in a population of $27$ retinal ganglion cells in the macaque retina \cite{pillow}. For a single neuron, the model assumes that the instantaneous spiking rate $r(t)$ is a nonlinear function $f$ of a sum of contributions,
\begin{equation}
r(t) = f\left[\vek{k}\cdot\vek{s}(t)  + \vek{q}\cdot\vek{y}(t_-) + \mu\right], \label{glm}
\end{equation}
where $\vek{k}$ is a linear filter acting on the stimulus $\vek{s}$, $\vek{q}$ is a linear filter acting on the spiking history $\vek{y}(t_-)$ of the neuron, and $\mu$ is an offset or an intrinsic bias towards spiking or silence. When the stimulus and the spike train are discretized into time bins of duration $\Delta$, the probability of observing (an integral number of) $y_t$ spikes is Poisson, with a mean given by $r_t \Delta$ (where the subscript indexes the time bin). Here, we neglect the history dependence of the spikes (with no loss of generality) and focus instead on the stimulus dependence; since each time bin is conditionally independent given the stimulus (and past spiking), the log likelihood for any spike train $\{y_t\}$ is \cite{pillow2}:
\begin{equation}
\log P(\{y_t\}|\vek{s}) = \sum_t y_t\log r_t - \Delta \sum_t r_t + c, \label{eqlike}
\end{equation}
where $c$ is independent of both $\mu$ and $\vek{k}$. This likelihood can be maximized with respect to $\mu$ and $\vek{k}$ (and optionally, with respect to $\vek{g}$) given adequate number of spikes, providing an estimate of  the filters from neural responses to complex, even natural stimuli. In contrast to maximally informative approaches, such as the stimulus energy derived in Section \ref{s2} \cite{Rajan+Bialek_12}, the functional form of the nonlinearity $f$ is an explicit assumption in likelihood-based methods like GLM. For specific classes of the function $f$, such as $f(z)=\log[1+\exp(z)]$, $\exp(z)$ or $[1+\exp(z)]^{-1}$, the likelihood optimization problem is convex and gradient ascent is guaranteed to find a unique global maximum. 

While the tractability consequent to convexity of the objective function is a big strength of this approach, the disadvantage is that if the chosen nonlinearity $f$ is different from the true function $f'$ used by the neuron, the filters inferred by maximizing likelihood  in Eq~(\ref{eqlike}) could be biased. If we relax the stringent requirement for convexity, we can choose more general nonlinear functions for the model, for example by parametrizing the nonlinearity in a point-wise fashion and inferring it jointly with the filters. For this discussion however, we assume that $f$ has been selected from the specific class of nonlinearities guaranteed to yield a convex likelihood function.

How can we extend GLM to situations where the neuron's response is more complex than a single linear projection of the stimulus? We will start with a proposal and follow up with a closely related formulation of~\citet{park+pillow_11} developed in parallel, which has provided a more complete analysis and several interesting extensions. One possibility is to expand the stimulus clip $\vek{s}$ of dimension $N$ into a larger space first, for instance by forming $\vek{s}\vek{s}^\tr$ (of dimension $N\times N$), and then operate on this object with a filter, i.e., $\sum_{i,j=1}^N (s_i s_j) Q_{ij}$. Such a term can be added to the argument of $f$ in the model exemplified in Eq~(\ref{glm}). Specifically, we propose a ``Generalized Quadratic Model" of the following form,
\begin{equation}
r(t) = f\left[\vek{k}\cdot\vek{s}(t)  + \vek{s}^\tr(t) \vek{Q}\vek{s}(t)+ \vek{g}\cdot\vek{y}(t_-) + \mu\right]. 
\end{equation}
If we want to retain convexity, we cannot expand $\vek{Q}$ in its eigenbasis and infer its vectors by maximizing the likelihood directly, because the eigenvectors appear quadratically. However, we can expand $\vek Q$ into a weighted sum of matrix basis functions, as in Eq~(\ref{basis}), making the argument of $f$ a linear function of basis coefficients $\alpha_\mu$,
\begin{equation}
r(t) = f\left(\vek{k}\cdot\vek{s}(t)  + \sum_{\mu=1}^M\left[\vek{s}^\tr(t) \vek{B^{(\mu)}}\vek{s}(t)\right] \alpha_\mu+ \vek{g}\cdot\vek{y}(t_-) + \mu\right).
\end{equation}
Existing methods for inferring GLM parameters \cite{pillow} can  be used to learn both the linear filter and the quadratic filter $\vek{Q}$ efficiently. After extracting $\vek{Q}$ we can check if a few principal components account for most of its structure (this is equivalent to checking whether $\vek{Q}$ is indeed a low rank matrix). In sum, this procedure provides a way of extracting multiple filters with GLM that is analogous to diagonalizing the spike-triggered covariance matrix on the Gaussian stimulus ensemble.

\begin{figure*}
\includegraphics[width=6in]{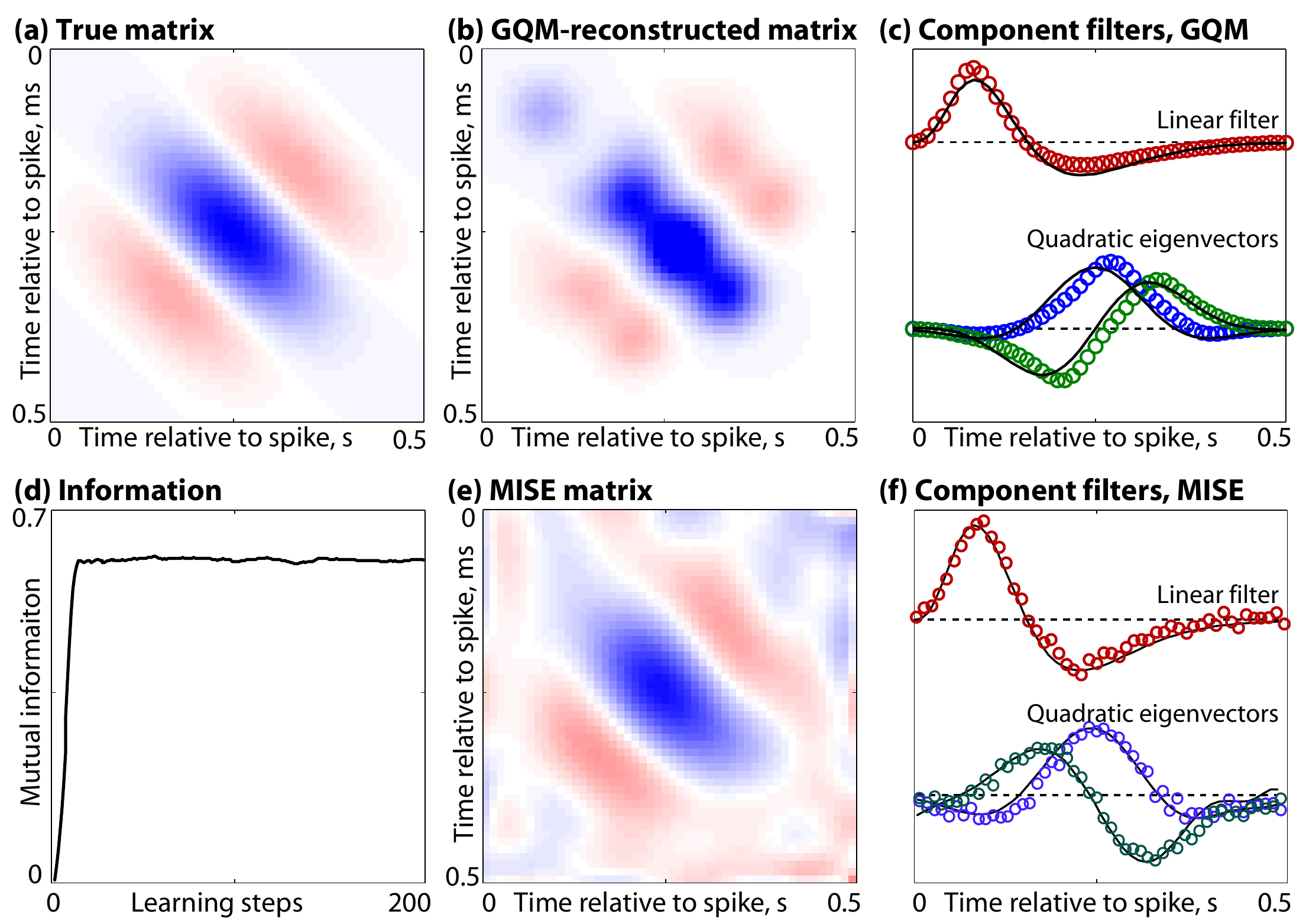}
\centering
\caption{{\bf Recovering the synthetic model of the contrast gain control cell using the flickering variance stimulus.} The spikes were simulated using the model presented in Fig.~\ref{fa}. {\bf (a)} The true quadratic kernel, $\mathbf{Q}$, of the model, is a matrix of rank 2 with the two filters combining into quadrature to estimate the signal ``power'' or variance. {\bf (b)} The reconstructed kernel using the quadratic extension of the GLM; the space of matrices was spanned by a 85-dimensional basis of Gaussian bumps (some of the granularity can still be seen in the reconstruction). The dominant eigenvectors of the inferred matrix are shown in {\bf (c)} in blue and green (solid black lines show the true values); shown is also the recovered linear filter (red circles) and its true value (solid black line). The inference of the same model using MISE shows quick convergence in {\bf (d)} and the recovered quadratic kernel in {\bf (e)}. {\bf (f)} The linear filter and the eigenvectors of the quadratic kernel recovered with MISE (circles), compared to the true values (black solid line). Note that quadratic filter eigenvectors are only determined up to a sign.}
\label{fb}
\end{figure*}

We have implemented such a quadratic extension to the GLM and applied it to the flickering variance stimulus shown in Fig.~\ref{fa}. The results are shown in Fig.~\ref{fb}a. The quadratic kernel correctly recovers a quadrature pair of filters;  we similarly recover the correct linear filter $\mathbf{k}_0$. While this method is restricted to a linear combination of first- and second-order filters within the nonlinearity, the distinct advantage over MISE is that the inference problem is convex with the appropriate nonlinearity.

\citet{park+pillow_11} consider an exponentiated general quadratic function of the following form (rewritten in the notation of this paper):
\begin{equation}
r(\mathbf{s}) = \exp\left(\mathbf{s}^\tr\mathbf{Q}\mathbf{s} + \mathbf{k}_0\cdot\mathbf{s} + \mu\right). \label{pp}
\end{equation}
First, the authors show that under a Gaussian stimulus ensemble, the expected log likelihood  can be expressed in terms of the STA, STC, and the covariance matrix of the stimulus, and derive the closed--form expressions for maximum likelihood estimates of  the quadratic kernel, linear filter, and the bias. Next, the generalization to arbitrary stimuli is achieved by numerically optimizing the true (as opposed to ``expected'') likelihood. In contrast to our suggestion of using the matrix basis expansion (which becomes an implicit regularizer upon choosing the dimensionality of the basis), Park and Pillow implement Bayesian regularization by imposing a prior on the quadratic kernel. This important suggestion is implemented as follows. 

The matrix is first decomposed into the eigensystem, $\mathbf{Q} = \sum_{i=1}^N \sigma_i\mathbf{w}_i\mathbf{w}^\tr_i$, where $\mathbf{w}$ are not forced to have an $L_2$ norm of 1, and $\sigma_i=\pm 1$ to indicate whether the filter $i$ is excitatory or suppresive (as in STC \cite{stnc}). Then, a zero-mean Gaussian prior $\mathcal{N}(0,\alpha_i^{-1}I)$ is put on each eigenvector $\mathbf{w}_i$, where the hyperparameter $\alpha_i$ determines the variance of the elements of eigenvector $i$; $\alpha_i\rightarrow\infty$ corresponds to eliminating the direction $i$ from the quadratic kernel and reducing its rank by 1. Next, an iterative algorithm is described for alternating between optimizing the likelihood with respect to model parameters, and optimizing the evidence given the parameters with respect to hyperparameters $\alpha_i$. This procedure correctly and efficiently identifies the rank of the quadratic kernels in synthetic examples, providing an automatic alternative for distinguishing ``significant'' from sampling-noise-induced eigenvectors in the STC and quadratic kernel inference. Finally, the authors show that Eq.~(\ref{pp}) can be further generalized at no additional computational cost from the exponentiated quadratic function  to a wider class of elliptic nonlinearities. 

To summarize, the reviewed work shows that the Bayesian generalization of STC and the generalization of GLMs to quadratic stimulus dependence yield equal probabilistic models for neural encoding that can be efficiently inferred for a restricted class of nonlinear functions. Attention needs to paid, however, to maintain the convexity of the procedure and deal with the large number of parameters in the quadratic kernel. To this end,  basis expansions as well as regularization with Bayesian priors seem like feasible candidates.
\section{Discussion} 
While powerful conceptually, the notion that neurons respond to multiple projections of the stimulus onto orthogonal filters is difficult to turn into a tractable inference procedure when the number of filters is larger than a few. To address this concern, alternative encoding models have recently been proposed where the neuron can be sensitive to higher-order features in the stimulus. Instead of being described by multiple linear filters, the neuron's sensitivity is described by a single quadratic filter (and optionally an additional linear filter). 
We have reviewed several inference methods for such quadratic stimulus dependence: two based on information maximization and the other based on maximizing the likelihood in an extension of generalized linear models. With MISE, no assumptions are made about how the projection onto the quadratic filter combines with the linear filter projection, and how both map into the probability of spiking. This approach yields unbiased filter estimates under any stimulus ensemble, but requires optimization in a possibly rugged information landscape. Noise entropy maximization is a flexible, maximum-entropy based framework for modeling the probability of spiking given stimulus.  It is computationally tractable and provides a convenient bound on the information per spike, but assumes a particular form of the nonlinearity. Alternatively, with a specific choice of nonlinearity and filter basis, likelihood inference within the GLM class can be extended to quadratic stimulus dependence while retaining the convexity of the objective function. By formulating the problem as Bayesian inference and choosing sparsifying priors for the quadratic filter, the true rank of the quadratic filter can also be inferred from data.  

All these approaches for inferring quadratic stimulus dependence are complementary; as we show in the appendix, maximum likelihood and information maximization inference also provide consistent filter estimates under defined conditions. A possible way to benefit from the tractability of likelihood formulations and maximization of noise entropy could be to use them to initialize a more general search using information maximization, in the hope that this would avoid the problems with the rugged information landscape, and remove the restrictions on the additive combination of linear and quadratic features.  

Examples of recent work establishing connections between  higher-order structure of natural scenes and neural mechanisms beyond the sensory periphery (e.g. \citet{karklin, tkacik}) make the development of corresponding methods for neural characterization, such as the ones presented here, very timely. Phenomena like phase invariance, adaptation to local contrast or sensitivity to signal envelope are widespread features of sensory neuron responses \cite{hubel, dan, baccus}. Moreover, as our abilities to record in vivo from the sensory systems of awake and behaving animals expand, so should the methods to analyze such recordings, where the relevant stimuli may no longer be perfectly controllable because of the animal's interaction with the environment \cite{kerr}.  The  methods presented here will help us systematically elucidate sensitivity to higher-order statistical features from responses of sensory neurons to natural stimuli.

\section*{Appendix: The relationship between information theoretic and likelihood-based inference}
We now demonstrate analytically that under rather general assumptions, the linear or quadratic filters obtained by maximizing mutual information match the filters inferred by maximizing the likelihood. We extend a reasoning we used previously in the context of inferring protein-DNA sequence-specific interactions in~\citet{kinney}, to neural responses. See also~\citet{Kouh} and references therein for a similar demonstration.

In the following, $x$ remains the projection of the stimulus $\vek{s}$ onto the linear ($x_t=\vek{k}\cdot\vek{s}_t$) or quadratic ($x_t=\vek{s}_t^\tr\vek{Q}\vek{s}_t$) filter, with time discretized in bins of duration $\Delta$ and indexed by subscript $t$. We collect all parameters that determine the filter into a vector $\vek{\theta}_1$. Given a single $x_t$, $y_t$ spikes are generated according to a conditional probability distribution $\pi(y_t|x_t)$. This probability distribution is typically assumed to be Poisson with mean given by $f(x_t)$ in the case of GLM, but we take a different approach. We discretize $x_t$ into $x=1,\dots,K$ bins and parameterize $\pi(y_t|x_t)$, which is a $Y_{\rm max} \times K$ matrix, by a set of parameters $\vek{\theta}_2$. Apart from assuming a cutoff value for the number of spikes per bin $Y_{\rm max}$ (which can always be chosen large enough to assign an arbitrarily low probability to observing  $>Y_{\rm max}$  spikes in any real dataset) and a particular discretization of the projection variable $x$, we leave the probabilistic relationship $\pi(y|x)$ between the projection and spike count completely unconstrained. The transformation from the stimulus to the spikes is then a Markov chain, fully specified by $\theta=\{\theta_1,\theta_2\}$,
\begin{equation}
\vek{s}_t \xrightarrow[\vek{k}\;\mathrm{or}\;\vek{Q}]{\vek{\theta}_1}x_t\xrightarrow[\pi]{\vek{\theta}_2}y_t.
\end{equation}
The likelihood of the spike train $\{y_t\}$ given the stimulus $\vek{s}$ is $P(\{y_t\}|\vek{s})=\prod_{t=1}^T \pi (y_t|x_t)$, where $T$ is the total number of time bins in the dataset. With $x$ discretized into $K$ bins, any dataset can be summarized by the count matrix $c_{yx}=\sum_{t=1}^T\delta(y,y_t)\delta(x,x_t)$, where $\delta$ is the Kronecker delta; note that $c_{yx}=T\tilde{p}(y,x)$, where $\tilde{p}$ is simply the empirical  distribution in the data of observing $y$ spikes jointly with projection $x$. In terms of $c$, the likelihood of the observed spike train is $P(\{y_t\}|\vek{s})=\prod_{y=0}^{Y_{\rm max}}\prod_{x=1}^K \pi (y|x)^{c_{yx}}$. Assuming that $x$ is adequately discretized and that $\pi$ is Poisson with mean $f(x)$, we will recover the generalized likelihood of Eq~(\ref{eqlike}).

Suppose that we are only interested in inferring the filter (parametrized by $\theta_1$), but not the filter-to-spike mapping $\pi$ (parameterized by $\theta_2$). While  avoiding any  assumptions about the structure of $\pi$, we can integrate the likelihood over $\theta_2$ with some prior $P_p(\theta_2$) such that
\begin{equation}
P(\{y_t\}|\vek{s})=\int d\theta_2\; P_p(\theta_2) \prod_{y,x}\pi(y|x)^{c_{yx}}. \label{newlike}
\end{equation} 
This resulting likelihood, called the \emph{model averaged likelihood}, is now only a function of $\theta_1$. The prior $P_p(\theta_2)$ can take many forms, but since we discretized $x$, thereby making $\pi(y|x)$ into  a (conditional probability) matrix, the simplest choice for the prior is  the \emph{uniform prior}. In this case we set $\theta_2$ equal to the entries in $\pi(y|x)$ matrix and  choose $P(\theta_2)$ to be uniform over all valid matrices $\pi$, such that the matrix entries are positive and the normalization constraint, $\sum_x \pi(y|x)=1$ for every $x$, is enforced. 

For any choice of priors we can rewrite Eq~(\ref{newlike}) as
\begin{equation}
P(\{y_t\} | \vek{s})=\int d\theta_2\; P_p(\theta_2)\exp\left[T\sum_{y,x}\tilde{p}(y,x)\log \pi(y|x)\right],
\end{equation}
which can be reorganized into
\begin{equation}
P(\{y_t\} | \vek{s})=\int d\theta_2 P_p(\theta_2)\exp\left[T\left\{\tilde{I}(y;x) - \tilde{S}(y) - \langle D_{KL}(\tilde{p}(y|x)\;||\;\pi(y|x))\rangle_{\tilde{p}(x)} \right\}\right].
\end{equation}
Here $\tilde{I}(y;x)=\sum_{y,x}\tilde{p}(y,x)\log\frac{\tilde{p}(y,x)}{\tilde{p}(y)\tilde{p}(x)}$ is the empirical mutual information between spike counts $y$ and the projection $x$, $\tilde{S}(y)$ is the empirical spike count entropy, and the ``correction'' term in brackets measures the average Kullback-Leibler divergence ($D_{KL}$) between the empirical and model conditional distributions. Importantly, only this correction term is a function of the $\pi$ and thus of $\theta_2$, and is  affected by the prior $P_p(\theta_2)$ which is being integrated over; the other terms can be pulled outside of the integral. We can therefore write the per time bin log likelihood as
\begin{equation}
\mathcal{L}=\frac{1}{T}\log P(\{y_t\} | \vek{s}) = \tilde{I}(y;x) - \tilde{S}(y) - \Lambda, \label{loglike}
\end{equation}
where the correction is
\begin{equation}
\Lambda=-\frac{1}{T}\log\int d\theta_2\; P_p(\theta_2) e^{-T\langle D_{KL}(\tilde{p}(y|x)\;||\;\pi(y|x))\rangle_{\tilde{p}(x)}}.
\end{equation}
It is necessary to show that as the amount of data $T$ grows, the correction $\Lambda$ decreases for a given choice of prior distribution $P_p(\theta_2)$, and for the choice of uniform prior this is  analytically tractable \cite{kinney}. Intuitively, it is clear that as $T\rightarrow\infty$, the empirical distribution $\tilde{p}(y|x)$ converges to the true underlying distribution $p(y|x)$, and the integral becomes dominated by the extremal point $\theta^*_2$, such that, in the saddle point approximation, 
\begin{equation}
\Lambda(T\rightarrow\infty)\sim \langle D_{KL}(p(y|x)\; ||\; \pi^*(y|x)) \rangle_{p(x)}.
\end{equation}
The distribution $\pi^*(y|x)$ is the closest distribution to $p(y|x)$ in the space over which the prior $P_p(\theta_2)$ is nonzero. As long as the prior assigns a non-zero probability to any (normalized) distribution, the divergence in $\Lambda$ will decrease and $\Lambda$ will vanish as $T$ grows. The case in which $\Lambda$ does not decay occurs when the prior completely excludes certain distributions by assigning zero probability, while the  data $p(y|x)$ precisely favors those excluded distributions.

Returning to the per time bin log likelihood $\mathcal{L}$ in Eq~(\ref{loglike}), as we decrease the time bin $\Delta$, we enter a regime where there is only 0 or 1 spike per bin, i.e., $y\in\{0,1\}$. Then the empirical information per time bin $\tilde{I}(y;x)$ can be written as,
\begin{equation}
\tilde{I}(y;x) = \tilde{p}(y=0) D_{KL}\left( \tilde{p}(x|y=0) || \tilde{p}(x) \right) + \tilde{p}(y=1)D_{KL}\left(  \tilde{p}(x|y=1) || \tilde{p}(x) \right),
\end{equation}
that is,
\begin{equation}
\tilde{I}(y;x) = \tilde{p}(\mathrm{silence}) \tilde{I}_{\rm silence} + \tilde{p}(\mathrm{spike})\tilde{I}_{\rm spike}.
\end{equation}

If the information in the spike train is dominated by the information carried in spikes \cite{brenner}, then the likelihood from Eq~(\ref{loglike}) becomes
\begin{equation}
\mathcal{L} = \tilde{p}(\mathrm{spike}) \tilde{I}_{\rm spike}+\dots, \label{equiv}
\end{equation}
where $\dots$ are terms that either do not depend of the filter parameters $\theta_1$ (i.e. entropy of the spike counts $\tilde{S}(y)$),  or vanish as the size of dataset grows ($\Lambda$). 

The identity in Eq~(\ref{equiv}) is the sought-after connection between the inference using information maximization and the likelihood-based approach. In the limit of small time-bins, maximizing the information per spike $I_{\rm spike}$ (in maximally informative approaches, as in \cite{mid} and Section~\ref{s2} of this paper),  on right-hand side of the identity, is the same as maximizing the \emph{model averaged likelihood} $\mathcal{L}$ of Eq~(\ref{loglike}), on the left-hand side of the identity.
\begin{acknowledgments}
We thank William Bialek and Michael J Berry II for insightful discussions and for providing critical scientific input during the course of this project. We would like to especially thank Jonathan Victor for helpful comments on the manuscript. This work was supported in part by the Human Frontiers Science Program, by the Swartz Foundation, by NSF Grants PHY--0957573 and CCF--0939370,  by the WM Keck Foundation, and by the ANR grant OPTIMA.

\end{acknowledgments}

\bibliography{template}
\end{document}